\documentstyle[prb,aps,twocolumn,epsfig]{revtex}
\begin{document}
\draft

\twocolumn[\hsize\textwidth\columnwidth\hsize\csname @twocolumnfalse\endcsname

\title{
Hidden non-Fermi liquid behavior due to crystal field quartet
}

\author{
Mikito Koga
}

\address{
Department of Physics, University of California Davis, CA 95616
}
\date{\today}
\maketitle

\begin{abstract}
We study a realistic Kondo model for crystal field quartet ground
states having magnetic and non-magnetic (quadrupolar) exchange couplings
with conduction electrons, using the numerical renormalization group method.
We focus on a local effect dependent on singlet excited states coupled
to the quartet, which reduces the non-magnetic coupling significantly and
drives non-Fermi liquid behavior observed in the calculated quadrupolar
susceptibility.
A crossover from the non-Fermi liquid state to the Fermi liquid state
is characterized by a small energy scale very sensitive to the non-magnetic
coupling.
On the other hand, the Kondo temperature observed in the magnetic
susceptibility is less sensitive.
The different crystal-field dependence of the two exchange couplings may
be related to the different $x$ dependence of quadrupolar and magnetic
ordering temperatures in Ce$_x$La$_{1-x}$B$_6$.
\end{abstract}


]

\section{Introduction}
The importance of orbital degrees of freedom has been emphasized
for heavy fermion systems such as Ce or U based compounds.
\cite{Cox98,Hewson93,Yamada92}
Since the $f$-electrons have localized character, the crystal-field
multiplets play important roles in the low temperature physics.
The multichannel Kondo effect due to the $f$-shell impurities is one of
the possible origins of the non-Fermi liquid (NFL) behavior in the heavy
fermion metals.\cite{Cox98}
The expected logarithmic divergence in their magnetic susceptibility $\chi$ and
specific heat coefficient $\gamma$ with decreasing temperatures is observed
in the dilute limit of the $f$-shell ions in U$_x$Th$_{1-x}$Ru$_2$Si$_2$
\cite{Ami94} or Ce$_x$La$_{1-x}$Pd$_2$Al$_3$.\cite{Nishi99}
This can be explained by the theory of the two-channel Kondo (2CK)
effect for each case.\cite{Cox93}
For the U case, the NFL behavior is due to a non-Kramers doublet in $f^2$,
while for the Ce case, it is associated with a Kramers doublet in $f^1$
coupled to a non-Kramers doublet in $f^2$.
These non-Kramers doublets give rise to two independent and equivalent
exchange interactions between the conduction electrons and the $f$-shell
moments.
The exchange coupling is quadrupolar for U, while it is magnetic
(dipolar or octupolar) for Ce.
We can also apply a similar scenario to account for a power-low
temperature dependent $\chi$ and $\gamma$ in UCu$_{5-x}$Pd$_x$.
\cite{Aron97,Andra98}
This NFL behavior may be caused by a quadrupolar Kondo coupling due to a
triplet ground state of U.\cite{Koga99}
\par

Our next interest in this work is what type of Kondo effect can occur for
a quartet $f$-electron state in realistic metals.
The quartet is coupled to the conduction electrons via a quadrupolar
exchange coupling as well as a dipolar one.
An octupolar exchange interaction has to be taken into account, too,
since it has comparable amplitude to the others.
The Kondo effect originating from the quartet is realized in
Ce$_x$La$_{1-x}$B$_6$, where the ground $\Gamma_8$ quartet is
separated very well from the excited $\Gamma_7$ doublet by 540K,
implying that the admixture of these states can be neglected at low
temperatures.
The observed $\chi$, $\gamma$, resistivity, and thermopower in this alloy
have been described by the Coqblin-Schrieffer (CS) Model with $SU(4)$
symmetry.\cite{Bick87}
This model is derived from an Anderson model for orbitally degenerate
conduction electrons and impurities where the valence fluctuation is
restricted to $f^0$-$f^1$ and the Coulomb interaction on the impurity
site is taken to be infinity.
In fact, the $SU(4)$ symmetry is broken in real systems because of
the $f^2$ configuration.
To confirm the robustness of the CS model, we have to take into
account more generalized exchange interactions associated with the atomic
structure.\cite{Hirst77}
Pavarini and Andreani studied the stability of $SU(N)$ symmetry against
some anisotropic exchange interactions based on their spherically symmetric
model.\cite{Pava97}
They concluded that the local Fermi-liquid (FL) fixed point derived
from the CS model is always stable and the Kondo temperature is reduced.
The recent study of Kusunose and Kuramoto on the same type of Kondo model
found that an NFL state is realized in the absence of non-magnetic exchange
couplings and is destabilized by the particle-hole symmetry breaking.
\cite{Kusu99}
\par

We cannot find any evidence of NFL behavior in the existing experimental
data for Ce$_x$La$_{1-x}$B$_6$.
It is not easy to obtain direct information of Ce single-site effects for
dense Ce cases since the long-range order occurs below 3.3K, very close to
the Kondo temperature ($\sim 2$K).
However, we note the difference between magnetic and quadrupolar effects
found in the phase diagram dependent on the Ce concentration $x$.
The transition temperature of the antiferro-quadrupolar order $T_{\rm Q}$
depends on $x$ more sensitively than the N{\'e}el temperature $T_{\rm N}$.
\cite{Hiroi97,Tayama97}
The former decreases very rapidly with decreasing $x$.
We can propose two possible scenarios to explain this behavior:
(1) the quadrupolar order is destabilized more easily by the lattice
disorder intensified with decreasing $x$, and (2) the quadrupolar exchange
coupling between the conduction electrons and the $\Gamma_8$ quartet becomes
smaller than the magnetic one, which results in the rapid decrease of the
quadrupolar RKKY coupling.
Since the latter has not been considered before, we stress that the strong
suppression of the quadrupolar exchange coupling is due to the crystal-field
states of Ce varying with decreasing $x$ from $x=1$.
It is also interesting that this quadrupolar property is very important for
the possibility of a new NFL state as discussed later.
\par

In this paper we present a realistic Kondo model with a spherical tensor
form.
By using a Schrieffer-Wolff transformation,\cite{Schrie66} it is derived
from an Anderson model with an orbitally degenerate impurity on which the
$f^1$ $\Gamma_8$ quartet is coupled to the $f^0$ configuration and the
$f^2$ $\Gamma_1$ singlet via the hybridization with the conduction electrons.
The model consists of a magnetic exchange term (dipolar and octupolar
interactions are combined) and a non-magnetic term (a potential scattering
and a quadrupolar exchange interaction are combined).
If we tune up the $f$-electron energy levels to leave the magnetic part alone,
we can obtain an NFL state corresponding exactly to that of the 2CK model,
except for the conversion of the charge and spin indices.
First, we discuss this NFL state realized only in a particle-hole symmetric
case, based on the numerical renormalization group (NRG) analysis.
\cite{Wilson75}
Next, we investigate the realistic case where the particle-hole symmetry
is broken and the effect of our NFL fixed point is observed in the
quadrupolar (strain) susceptibility.
The most important result is the existence of two characteristic energy
scales:
one corresponds to the Kondo temperature associated with the CS limit, while
the other characterizes a crossover to the FL state at lower temperatures.
The latter new scale is connected to the NFL fixed point and can be very
small when the quadrupolar exchange coupling is small.
This quadrupolar coupling destabilizes the NFL state in the same manner as
a spin field in the 2CK model.\cite{Affl92}
\par

The remainder of this paper is organized as follows.
In Sec.~II, we explain the different crystal-field dependence of the magnetic
and non-magnetic exchange couplings and describe how to calculate magnetic
and quadrupolar susceptibilities for the impurity with the NRG method.
In Sec.~III, we discuss the FL and NFL states based on the NRG results.
Finally, Sec.~IV gives the concluding remarks and discussions with some
proposals for experiments.
\par

\section{Model}
As mentioned above, we restrict the $f$-electron states to $f^0$,
an $f^1$ $\Gamma_8$ quartet, and an $f^2$ $\Gamma_1$ singlet in our
Anderson model.
These are possible ground states, accessed by varying two parameters of a
crystal-field Hamiltonian.\cite{Lea62}
Here we assume that the energies of the excited states in each
configuration are so large as to be neglected at low temperatures.
Only in this case can we map a derived Kondo model to an exchange
model expressed with a spherical tensor form.
Applying a Schrieffer-Wolff transformation to the $f^0$-$f^1$ valence
fluctuation, we obtain the CS model with $SU(4)$ symmetry.
To rewrite this Hamiltonian with spherical tensor operators, we express the
fourfold $\Gamma_8$ states in terms of $J_z$ of a $J=3/2$ spin operator in the
following way:\cite{Onode66}
\begin{eqnarray}
& & |\Gamma_{8,3/2} \rangle =
\mbox{} -\sqrt{1 \over 6}|+{3 \over 2} \rangle
\mbox{} -\sqrt{5 \over 6}|-{5 \over 2} \rangle,
\label{eq:model1} \\
& & |\Gamma_{8,1/2} \rangle = |+{1 \over 2} \rangle, \\
& & |\Gamma_{8,-1/2} \rangle = -|-{1 \over 2} \rangle, \\
& & |\Gamma_{8,-3/2} \rangle =
\mbox{} \sqrt{1 \over 6}|-{3 \over 2} \rangle
\mbox{} +\sqrt{5 \over 6}|+{5 \over 2} \rangle,
\label{eq:model2}
\end{eqnarray}
where we assume that these states are constructed from the single-electron
states with the total angular momentum $j=5/2$ because of a strong
spin-orbit coupling in Ce, and $|M \rangle$ ($M = -5/2,-3/2,\cdots,5/2$)
represents an eigenstate of $j_z$.
The CS Hamiltonian can be expanded with respect to scalar products of two
kinds of irreducible tensor operators with rank $p$, and the corresponding
effective exchange Hamiltonian is given by
\begin{eqnarray}
& & H_{{\rm ex},10} =
I_{10} \sum_{kk'mm'} a_{k'm'}^{\dag} a_{km}
\nonumber \\
& &~~~~~~~~
\times \left[{1 \over 5} (T_{m'm}^{(1)} + {4 \over 9} T_{m'm}^{(3)})
\mbox{} +{1 \over 4}(1 + {4 \over 9} T_{m'm}^{(2)}) \right],
\label{eq:model3}
\end{eqnarray}
where
\begin{equation}
T_{m'm}^{(p)} = \sum_{q=-p}^p (-1)^q
\left(j_{-q}^{(p)} \right)_{m'm} J_q^{(p)}
\end{equation}
represents a dipolar, a quadrupolar, and an octupolar exchange interaction
for $p=1$, $p=2$, and $p=3$, respectively.
Here $a_{km}^{\dag}$ ($a_{km}$) is a creation (annihilation) operator
for a conduction electron with wave vector $k$ and orbital $m$.
The tensor $j_q^{(p)}$ ($J_q^{(p)}$) is constructed from a $j=3/2$ ($J=3/2$)
spin operator for the fourfold $\Gamma_8$ conduction electrons
(the $\Gamma_8$ quartet).
For each $p$, the tensor operator has $(2p+1)$ components and is derived
from the following equations,
\begin{eqnarray}
& & J_p^{(p)} = (-1)^p \sqrt{(2p-1)(2p-3) \cdots 3 \cdot 1 \over
2p(2p-1) \cdots 2}J_+^p, \\
& & J_{q-1}^{(p)} = {1 \over \sqrt{(p+q)(p-q+1)}} [J_-, J_q^{(p)}].
\end{eqnarray}
We take the Fermi energy as the origin of energies, and the coupling
constant $I_{10}$ is given by $V_{10}^2 / \Delta E_{10}$, where $V_{10}$ is
the mixing parameter for $f^0$-$f^1$ and $\Delta E_{10}$ is the energy level
of $f^0$ measured from that of the $f^1$ $\Gamma_8$ quartet.
The valence fluctuation from $f^1$-$f^2$ also leads to the similar form of the
effective exchange Hamiltonian since the twofold $\Gamma_7$ ($j=5/2$)
conduction electrons cannot contribute to this exchange process, being
restricted by the $\Gamma_1$ symmetry
($\Gamma_1 = \Gamma_6 \otimes \Gamma_6 \oplus \Gamma_7 \otimes \Gamma_7
\oplus \Gamma_8 \otimes \Gamma_8$)
of the intermediate state in $f^2$.
The SU(4) symmetry is also conserved in the exchange interaction, and we
obtain the corresponding Hamiltonian for $f^1$-$f^2$,
\begin{eqnarray}
& & H_{{\rm ex},12} =
I_{12} \sum_{kk'mm'} a_{k'm'}^{\dag} a_{km}
\nonumber \\
& &~~~~~~~~
\times \left[{1 \over 5} (T_{m'm}^{(1)}+{4 \over 9} T_{m'm}^{(3)})
\mbox{} -{1 \over 4}(1 + {4 \over 9} T_{m'm}^{(2)}) \right],
\label{eq:model4}
\end{eqnarray}
where the sign of the potential scattering and quadrupolar coupling terms is
negative, while it is positive in the Hamiltonian (\ref{eq:model3}).
Here the coupling constant $I_{12}$ is given by
$V_{12}^2 / \Delta E_{12}$, where the definition of each parameter
follows that for $I_{10}$.
If the $\Gamma_8 \otimes \Gamma_8$ part of the $f^2$ $\Gamma_1$ state
consists of single-electron states given in
(\ref{eq:model1})--(\ref{eq:model2}), we can obtain
$V_{12} / V_{10} = 1 / \sqrt{6}$ which is related to the Clebsch-Gordan
coefficients.
After we put the two Hamiltonians (\ref{eq:model3}) and
(\ref{eq:model4}) together, we obtain the complete exchange Hamiltonian
with two exchange couplings as follows:
\begin{equation}
H_{\rm ex} = \sum_{kk'mm'} a_{k'm'}^{\dag} a_{km}
(I_{\rm o} T_{m'm}^{({\rm o})} + I_{\rm e} T_{m'm}^{({\rm e})}),
\label{eq:model5}
\end{equation}
where
\begin{eqnarray}
& & T_{m'm}^{({\rm o})} = {1 \over 5} (T_{m'm}^{(1)}
\mbox{} + {4 \over 9} T_{m'm}^{(3)}), \\
& & T_{m'm}^{({\rm e})} = {1 \over 4} (1 + {4 \over 9} T_{m'm}^{(2)}), \\
& &{I_{\rm e} \over I_{\rm o}} =
{1 - (I_{12} / I_{10}) \over 1 + (I_{12} / I_{10})}~~(\equiv \alpha).
\label{eq:model6}
\end{eqnarray}
Here $I_{\rm o}$ is always positive and antiferromagnetic, while the
sign of $I_{\rm e}$ can change.
We note that the transformation $I_{\rm e} \leftrightarrow -I_{\rm e}$ only
causes the conversion between particles and holes.
The ratio of the couplings $\alpha$ represents the anisotropy apart
from the $SU(4)$ symmetry ($\alpha = 1$).
As shown in Eq.~(\ref{eq:model6}), the quadrupolar coupling can be very small
depending on the exchange couplings $I_{10}$ and $I_{12}$, i.e., on the
valence fluctuations $f^0$-$f^1$ and $f^1$-$f^2$.
Thus the quadrupolar coupling is suppressed by the local environment more
strongly than the dipolar and octupolar couplings.
Recently Kusunose and Kuramoto studied the same type of Kondo model for
arbitrary ranks of tensors by the perturbation renormalization group analysis.
\cite{Kusu99}
They found that an unstable NFL fixed point is realized when all the
exchange interactions expressed by $j_{-q}^{(2n)} J_q^{(2n)}$ ($n$ is integer)
vanish and particle-hole symmetry is conserved.
This corresponds to the case $I_{\rm e} = 0$ ($I_{10} = I_{12}$) in our
realistic Kondo model.
This is realized  when $\Delta E_{12} / \Delta E_{10}$ is exactly
equal to $(V_{12} / V_{10})^2$ ($= 1/6$) for the crystal-field energies
of $f$-electrons.
Such complete cancellation of $I_{10}$ and $I_{12}$ is unlikely in real
systems, but the effect of the NFL state should be observed at finite
temperatures as discussed later.
\par

To investigate our Kondo model given by
\begin{equation}
H = \sum_{km} \varepsilon_k a_{km}^{\dag} a_{km} + H_{\rm ex},
\end{equation}
where $\varepsilon_k$ is the kinetic energy of the conduction electron,
we transform the Hamiltonian to a hopping Hamiltonian for the NRG calculation.
Following Wilson,\cite{Wilson75} we obtain the recursion relation
\begin{equation}
H_{N+1} = \Lambda^{1/2} H_N + (f_{N+1,m}^{\dag} f_{Nm} + {\rm H.c.}),
\end{equation}
and
\begin{equation}
H_0 = \Lambda^{-1/2} \sum_{mm'} f_{0,m'}^{\dag} f_{0,m}
(\tilde{I}_{\rm o} T_{m'm}^{({\rm o})}
\mbox{} + \tilde{I}_{\rm e} T_{m'm}^{({\rm e})}),
\end{equation}
for the impurity, where through the Fourier transformation $f_{nm}^{\dag}$
($f_{nm}$) is derived from $a_{km}^{\dag}$ ($a_{km}$) defined in the
logarithmically descretized conduction band, and $\tilde{I}_{\rm o (e)}$
is equal to $2 I_{{\rm o} ({\rm e})} \rho / (1 + \Lambda^{-1})$, where
$\rho$ is the density of states of the conduction electrons at the Fermi sea.
In order to obtain the temperature dependent magnetic ($\chi$) and
quadrupolar ($\chi_{\rm Q}$) susceptibilities for the local $\Gamma_8$
quartet, we calculate the magnetization $M_z$ and quadrupolar response
$Q_{zz}$, respectively, to very small local external fields.
For the latter, the lattice is distorted along a c-axis.
In the NRG calculation, $M_z$ and $Q_{zz}$ are given by
\begin{eqnarray}
& & M_z = {{\rm Tr} J_z \exp (-\bar{\beta} H'_N)
\over {\rm Tr} \exp (-\bar{\beta} H'_N)},
\label{eq:model7} \\
& & Q_{zz} = {{\rm Tr} (\lambda O_0^2) \exp (-\bar{\beta} H''_N)
\over {\rm Tr} \exp (-\bar{\beta} H''_N)}.
\label{eq:model8}
\end{eqnarray}
Here $O_0^2$ is defined by $[3 J^2 - J(J+1)]$ and $\bar{\beta}$ ($\sim 1$) is
related to the physical temperature
\begin{equation}
k_{\rm B}T/D = {1 + \Lambda^{-1} \over 2} \Lambda^{-(N-1)/2} / \bar{\beta},
\end{equation}
where $D$ is a half width of the conduction band.
For the Hamiltonian $H'_N$ in Eq.~(\ref{eq:model7}), the Zeeman term
$-g_J \mu_{\rm B} J_z h_z$ ($g_J = 6/7$ is a Land{\'e}'s $g$ factor for the
$\Gamma_8$ quartet and $h_z$ is a local magnetic field) is added to the
exchange Hamiltonian $H_0$.
For $H''_N$ in Eq.~(\ref{eq:model8}), the similar term $-\lambda O_0^2 h_q$
($h_q$ is a local quadrupolar field) is added to $H_0$, where we determine
$\lambda$ to satisfy
\begin{eqnarray}
& & \langle \Gamma_{8,\pm 3/2} | \lambda O_0^2 | \Gamma_{8,\pm 3/2} \rangle
= +1, \\
& & \langle \Gamma_{8,\pm 1/2} | \lambda O_0^2 | \Gamma_{8,\pm 1/2} \rangle
= -1.
\end{eqnarray}
The susceptibilities $\chi$ and $\chi_{\rm Q}$ are given by $M_z / h_z$ and
$Q_{zz} / h_q$, respectively, if the local external fields are small enough.
Throughout this paper we take $k_{\rm B}$, $\mu_{\rm B}$ and $D$ to be unity.
In the NRG calculation, we take $\Lambda = 3$ and keep about 800 lowest-lying
states at each renormalization step.
\par

\section{Results}
Let us begin with the case $\alpha = 0$ in (\ref{eq:model6}).
Because of the particle-hole symmetry, we can express eigenstates with an
axial charge operator $\vec{q}$ defined by \cite{Kim97}
\begin{eqnarray}
& & q_+ = \sum_{n=0}^{\infty}
(f_{n,3/2}^{\dag} f_{n,-3/2}^{\dag} - f_{n,1/2}^{\dag} f_{n,-1/2}^{\dag}), \\
& & q_z = {1 \over 2} \sum_{n=0}^{\infty} \sum_m
(f_{nm}^{\dag} f_{nm} - {1 \over 2}).
\end{eqnarray}
This operator satisfies the $SU(2)$ Lie algebra,
\begin{equation}
[q_z,q_{\pm}] = \pm q_{\pm},~~[q_+,q_-] = 2q_z.
\end{equation}
After a large number of renormalization steps $N$, we reach an NFL fixed point
where $H_N = H_{N+2}$ is satisfied.
Each energy level is labeled by $(q,j)$ where $j$ is the total angular
momentum.
As shown in Table~I (a), the ground state has twofold degeneracy related to
the particle-hole symmetry ($q_z = \pm 1/2$) and no angular momentum
($j = 0$).
The lowest energy spectrum is same as that of NFL fixed point for the 2CK
model.
The eigenstates for the latter are given in (b).
The energies in Table~I are derived by the boundary conformal field theory
(CFT) \cite{Affl90} which is applied to the 2CK model with the
$U(1) \times SU(2) \times SU(2)$ symmetry.
Here $v_{\rm F}$ is the Fermi velocity and $l$ is the system size.
Each energy is equal to the corresponding NRG energy multiplied by a factor
($\simeq 5/8$) for $\Lambda = 3$ case.\cite{Affl92}
Our Kondo model with particle-hole symmetry has the $SU(2)$ symmetry
for both $q$ and $j$, and each of the four lowest-lying eigenstates in (a)
satisfies the following equation
\begin{equation}
E = {v_{\rm F} \pi \over l}
\left[{q(q+1) \over 4} + {j(j+1) \over 12} \right],
\end{equation}
for the corresponding energy.
Each energy in the last column is measured from that of the ground state
($= 3/16$).
These eigenstates can be connected to the primary states in the CFT.
The CFT gives the same equation to NFL fixed-point energies for a Kondo model
with only dipolar exchange interaction between $j=3/2$ conduction
electrons and a $J=1/2$ local spin.\cite{Kim97}
This indicates applicability of the CFT to our Kondo model with the octupolar
exchange interaction, although it is impossible to absorb the impurity spin
in the conduction electron current.
\par

The new NFL fixed point is stabilized by the octupolar exchange interaction
$T^{(3)}$ in the Hamiltonian (\ref{eq:model5}).
On the other hand, the dipolar exchange interaction $T^{(1)}$ leads to
another NFL fixed point if both $T^{(2)}$ and $T^{(3)}$ are neglected.
The lowest NRG energy levels in this case are same as those for the Kondo model
with exchange interaction between $j=3/2$ conduction electrons and a
$J=1/2$ local spin, if even and odd numbers of the NRG iteration step $N$
are converted each other.
The NFL fixed point due to $T^{(1)}$ is stable against the potential
scattering and is destabilized by $T^{(2)}$ and $T^{(3)}$.
If only $T^{(1)}$ and $T^{(3)}$ are left in the exchange interaction and
these couplings are varied independently, we can obtain the competition
between each other.
For the exchange Hamiltonian (\ref{eq:model5}), we always reach the
NFL fixed point stabilized by $T^{(3)}$ when we fix $\alpha =0$.
\par

The particle-hole symmetry is usually broken ($\alpha \neq 0$) and the
quadrupolar coupling exists in any real systems.
In Table~I, the axial charge $q$ for the new NFL fixed point (a) completely
corresponds to the spin $S$ for the 2CK fixed point (b).
As a spin field in the 2CK model, the potential scattering and
quadrupolar exchange terms destabilize the new NFL fixed point, and leads to
the $SU(4)$ symmetric FL fixed point realized for the CS model.
The crossover from the NFL state to the FL state is related to the recovery
of the $SU(4)$ symmetry.
In the NRG calculation, the doublet ground state ($q = 1/2, j = 0$) for
$\alpha = 0$ is lifted by the coupling $I_{\rm e}$.
The NRG energy flow diagram shows that one of the separated states becomes
unstable as the renormalization step $N$ increases, and finally merges to a
fivefold degenerate excited state with $j=2$.
We define the crossover temperature $T_{\rm s}^*$ as the number of the NRG
step at which the energy difference between the two states reach less than
$0.01$.
Figure~1 shows that $T_{\rm s}^*$ is proportional to $\alpha^2$, similar to
a crossover temperature vs. a spin field for the 2CK model.
\cite{Affl92}
\par

Next we discuss the crossover to the FL state by calculating $\chi(T)$ and
$\chi_{\rm Q}(T)$.
In the CS limit ($\alpha =1$), these susceptibilities show the same
temperature dependence.
If they are normalized by the $T=0$ values, we can obtain the complete
agreement with each other.
The difference is enhanced as the symmetry of the exchange interaction
is lowered from the $SU(4)$ symmetry, i.e, $\alpha$ decreases from unity.
In Fig.~2, $\chi_{\rm Q} (T)$ for $\alpha = 0$ behaves logarithmically
down to $T = 0$ as does the spin susceptibility for the 2CK model.
Based on this logarithmic curve, we can estimate the Kondo temperature
$T_{\rm K}^{(0)}$ as a function of the magnetic coupling $I_{\rm o}$.
Once the non-magnetic coupling $I_{\rm e}$ is introduced ($\alpha \neq 0$),
the FL state appears below $T_{\rm s}^*$ and $\chi_{\rm Q}$ becomes constant.
The $T=0$ value of $\chi_{\rm Q}$ decreases as $\alpha$ increases.
As shown in Fig.~3, $\chi_{\rm Q} (T=0) \propto \ln \alpha$ is satisfied
for the small values of $\alpha$.
In analogy with the spin susceptibility for the two-channel Kondo model with
channel anisotropy,\cite{Fabri95,Andrei95} $\chi_{\rm Q}$ is scaled by
$T_{\rm K}^{(0)}$ and $T_{\rm s}$:
the latter is proportional to $T_{\rm s}^*$.
Since $T_{\rm s}$ is proportional to $\alpha^2$, we can express $\chi_{\rm Q}$
with the following equation
\begin{equation}
\chi_{\rm Q} = {a \over T_{\rm K}^{(0)}}
\ln \left[{T_{\rm K}^{(0)} \over {\rm max}(T_{\rm s},T)} \right],
\end{equation}
for $T_{\rm s} \ll T_{\rm K}^{(0)}$.
We obtain $(a,T_{\rm K}^{(0)},T_{\rm s} / \alpha^2) =
(0.60, 2.6 \times 10^{-3}, 8.1 \times 10^{-3})$ for $\tilde{I}_{\rm o} = 0.2$
and $(0.62, 3.1 \times 10^{-2}, 6.1 \times 10^{-2})$ for
$\tilde{I}_{\rm o} = 0.3$.
\par

These two energy scales are also found in $\chi$, which we conclude from the
detailed analysis shown below.
For the CS model with the exchange interaction between $j=3/2$ conduction
electrons and a $J=3/2$ local spin, the Kondo temperature $T_{\rm K}^{(1)}$
is defined by $j(j+1) / 3 \chi (T=0)$.
The $\chi(T) / \chi(0)$ ($\alpha = 1$) curve for the $\Gamma_8$ quartet is
in agreement with that for the $J=3/2$ spin completely, and we obtain
$\chi(T_{\rm K}^{(1)}) / \chi(0) \simeq 1/3$.
As shown in Fig.~4, a crossover to the FL state appears with a $\ln T$
dependence in $\chi(T)$ for each value of $\alpha$.
We also obtain $\chi(T_{\rm K}^{(0)}) / \chi(0) \simeq 1/3$ for $\alpha = 0$,
where $T_{\rm K}^{(0)}$ has been obtained from the logarithmic temperature
dependent $\chi_{\rm Q}$ for $\alpha = 0$.
Therefore we can define the Kondo temperature for arbitrary values of $\alpha$
using the equation $\chi(T_{\rm K}^{(\alpha)}) / \chi(0) = 1/3$.
Based on this analysis, we find that $T_{\rm K}^{(\alpha)}$ is analogous to
$T_{\rm K}^{(1)}$ of the CS limit when $|\alpha|$ is close to unity.
In Fig.~5, $\chi (0)$ is fitted well to the function
$(T_{\rm K}^{(0)} + b \alpha^2)^{-1}$ from $\alpha = 1$ down to
$\alpha \simeq 0.2$ for $\tilde{I}_{\rm o} = 0.2$ ($\alpha \simeq 0.4$ for
$\tilde{I}_{\rm o} = 0.3$), where the constant $b$ depends on $I_{\rm o}$.
On the other hand, we find that $T_{\rm K}^{(\alpha)}$ is proportional
to the inverse of this function in the same parameter region of $\alpha$, and
finally obtain
\begin{equation}
T_{\rm K}^{(\alpha)} \simeq 1.3 (T_{\rm K}^{(0)} + c T_{\rm s})
\simeq {1.3 \over \chi(0)},
\end{equation}
where $c = 1.4$ for $\tilde{I}_{\rm o} = 0.2$ and $c = 0.86$ for
$\tilde{I}_{\rm o} = 0.3$.
This relation implies that the CS model can describe $\chi$ very well even
if the $SU(4)$ symmetry is broken in real systems, since $|\alpha|$ is not
usually so small.
The strength of the anisotropy ($|\alpha| < 1$) is related to the reduction
of the Kondo temperature.
The large decrease of the quadrupolar coupling is seen in $\chi$ as a
small difference from the CS limit.
On the other hand, the crossover depending on the quadrupolar coupling can be
observed more easily in $\chi_{\rm Q}$, since the observable crossover
temperature $T_{\rm s}$ is very sensitive to the quadrupolar coupling.
\par

\section{Discussions}
We have discussed the multipolar Kondo effect due to the $\Gamma_8$ quartet,
based on the realistic Kondo model with the spherical tensor form.
The exchange couplings depend on the crystal-field energy levels of the $f^0$,
$f^1$ $\Gamma_8$ quartet, and $f^2$ $\Gamma_1$ singlet states.
We can tune up the crystal field to leave the magnetic (dipolar $+$ octupolar)
exchange coupling $I_{\rm o}$ alone and obtain the new NFL state at low
temperatures in this particle-hole symmetric case.
This symmetry is usually broken by the non-magnetic
(quadrupolar $+$ potential) coupling $I_{\rm e}$ in real systems.
The latter coupling $I_{\rm e}$ behaves like a spin field in the 2CK model.
The crossover from the NFL state to the FL state is related to the ratio of
the two couplings $\alpha$ and we obtain two characteristic energy scales:
one is the Kondo temperature $T_{\rm K}^{(\alpha)}$ reduced by the decrease
of $\alpha$ from the CS limit ($\alpha = 1$), and the other is the small
crossover temperature $T_{\rm s}$ ($\propto \alpha^2$) due to the NFL fixed
point.
The latter energy scale $T_{\rm s}$ clearly appears with the suppression of
the logarithmic behavior in the quadrupolar susceptibility $\chi_{\rm Q} (T)$.
\par

In our Kondo model, we have neglected the effect of the local excited states
($\Gamma_3$, $\Gamma_4$, $\Gamma_5$) in $f^2$, assuming that their energies
are very large.
If they are included, the Kondo model cannot be expressed with the spherical
tensor form.
The exchange process via the excited states makes the exchange interaction
much more complicated, and prevents the total angular momentum from being
a good quantum number if $\Gamma_8$ is mapped to $J=3/2$.
In addition, the twofold $\Gamma_7$ conduction electrons contribute to the
exchange with the $\Gamma_8$ quartet and leads to a more complicated Kondo
interaction.
The larger number of degrees of freedom of the scattering process may stabilize
the FL state more.
For Ce$_x$La$_{1-x}$B$_6$, we can expect a large crystal-field splitting
between the ground and excited states in $f^2$ as well as between the
$\Gamma_8$ and $\Gamma_7$ states in $f^1$.
Since the $f$-electron is localized very well (the $f$-occupancy is close
to one), the hybridization of the conduction band with the $f$-orbitals
may be so small that the effect of the excited states becomes irrelevant.
However, it is not clear which is a ground state for $f^2$ among the
possible candidates $\Gamma_1$, $\Gamma_3$, and $\Gamma_5$.
\par

Finally we discuss the connection between our results and the experiments
for Ce$_x$La$_{1-x}$B$_6$.
According to the magnetic susceptibility and the resistivity, the Kondo
temperature $T_{\rm K}$ is not sensitive to the Ce concentration $x$.
\cite{Sato84,Sato85}
In addition, the $x$ dependence of the lattice parameter is very small.
\cite{Sato84}
In our theory, $T_{\rm K}$ depends on the two Kondo couplings $I_{\rm o}$
and $I_{\rm e}$.
As we discussed in the introduction, $T_{\rm Q}$ decreases more rapidly than
$T_{\rm N}$ with decreasing $x$ from $x=1$.
One can usually expect the quadrupolar order to be much sensitive to
the lattice disorder since the quadrupolar moment is directly coupled to the
lattice degrees of freedom.
However, doping also changes the crystal field.
It is possible that the quadrupolar RKKY coupling is suppressed more strongly
by the change of $x$ than the magnetic one, due to the different crystal-field
dependence of the quadrupolar and magnetic exchange couplings of the
conduction electrons with the $\Gamma_8$ quartets.
We can conclude that the ratio of the two couplings
$|I_{\rm e} / I_{\rm o}|$ decreases as $x$ decreases from $x=1$.
Based on our results, $T_K$ decreases with $|I_{\rm e} / I_{\rm o}|$, while
it increases if $I_{\rm o}$ becomes larger.
This is possible in general and thus we can obtain $T_{\rm K}$ to be less
sensitive to doping than $T_{\rm Q}$ even if the Kondo couplings are varied.
\par

In order to confirm our idea for the $x$ dependence of $T_{\rm Q}$, we
propose that the experimentalists measure $\chi$ and $\chi_{\rm Q}$ together
for arbitrary $x$ including the dilute limit ($x \ll 1$).
$\chi_{\rm Q}$ is related to the elastic constant $(C_{11} - C_{12}) / 2$.
It is very useful to observe the correlation between the two susceptibilities
for the purpose of finding the local effect.
When $|I_{\rm e} / I_{\rm o}| \simeq 1$, the correlation is expected to be
linear in a temperature region where the Kondo effect is observed, since
both $\chi$ and $\chi_{\rm Q}$ show almost the same temperature dependence.
This linear feature tends to be lost as $|I_{\rm e} / I_{\rm o}|$ decreases
form unity.
Large deviations from the linear correlation can be found at low temperatures
below $T_{\rm K}$.
At higher temperatures, slight deviations are observable when $|I_{\rm e}|$ is
very small.
The same argument can be applied to the $C_{44}$ mode as well.
The amplitude of the quadrupolar Kondo coupling is also related to the type
of quasiparticle excitations, particle-like or hole-like.
If we are able to extract the single impurity effect, the thermopower
measurement can detect the degree of particle-hole asymmetry which becomes
smaller as the quadrupolar coupling decreases.
It is also very interesting to search for the hidden NFL state related to
the particle-hole symmetry not only in Ce$_x$La$_{1-x}$B$_6$ but also
in other cubic crystals including rare earth ions where a $\Gamma_8$
quartet is a ground state.
\par

\acknowledgements
The author thanks D. L. Cox and R. Shiina for stimulating discussions
and critical reading of this manuscript.
This research has been supported by JSPS Postdoctoral Fellowships for
Research Abroad and by the US Department of Energy, Office of Science,
Division of Materials Research.
\par

\begin{table}
\begin{displaymath}
\begin{array}{cccccccccc}  \hline
 \\
({\rm a}) & q & j & & ({\rm b}) & Q & S & S_{\rm c} & ~~~
 & El / \pi v_{\rm F} \\
 \\ \hline
 \\
 & 1/2 & 0 & & & 0 & 1/2 & 0 & & 0 \\
 \\
 & 0 & 3/2 & & & \pm 1 & 0 & 1/2 & & 1/8 \\
 \\
 & 1/2 & 2 & & & \pm 2 & 1/2 & 0 & & 1/2 \\
 \\
 & & & & & 0 & 1/2 & 1 & & 1/2 \\
 \\
 & 1 & 3/2 & & & \pm 1 & 1 & 1/2 & & 5/8 \\
 \\
 & 3/2 & 0 & & & 0 & 3/2 & 0 & & 1 \\
 \\
 & 1/2 & 3 & & & \pm 2 & 1/2 & 1 & & 1 \\
 \\
 & & & & & 0 & 1/2 & 0 & & 1 \\
 \\
 & 1/2 & 1 & & & 0 & 1/2 & 1 & & 1 \\
 \\
 & 1/2 & 0 & & & 0 & 1/2 & 0 & & 1 \\
 \\ \hline
\end{array}
\end{displaymath}
\vspace*{0.05cm}
\caption{
Lowest energy spectra at non-Fermi liquid fixed points:
(a) For our model ($\alpha = 0$), $q$ is an axial charge and $j$ is a total
angular momentum.
(b) For the two-channel Kondo model, $Q$ is a total number of particles
measured from that of the ground state, $S$ is a total spin, and $S_{\rm c}$
is an isospin associated with the two channels.
The last column gives energies corresponding to the eigenstates for both (a)
and (b), where we take the energy of the ground state as the origin.
}
\end{table}
\begin{figure}
\begin{center}
\psfig{figure=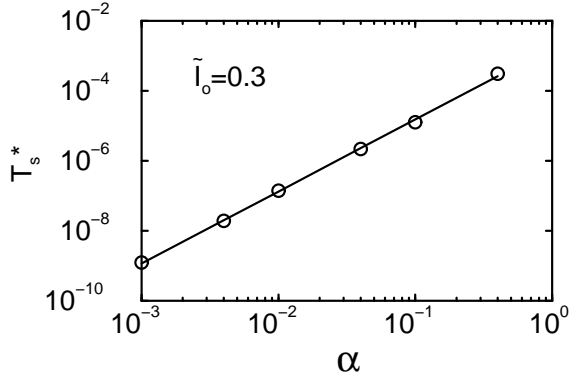,width=7.5cm}
\end{center}
\vspace*{0.05cm}
\caption{
Crossover temperature.
Over a wide range of the coupling ratio $\alpha$, $T_{\rm s}^*$ is
proportional to $\alpha^2$.
}
\end{figure}
\begin{figure}
\begin{center}
\psfig{figure=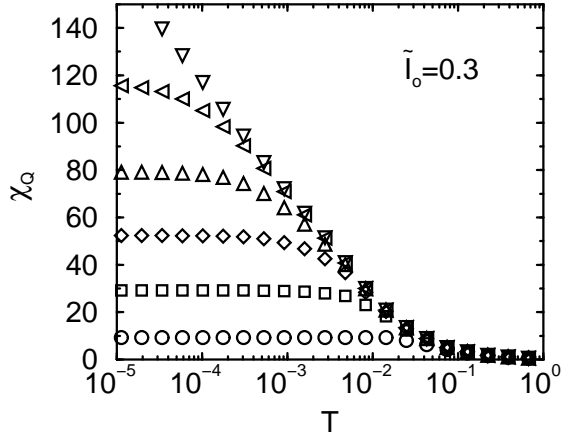,width=7.5cm}
\end{center}
\vspace*{0.05cm}
\caption{
Temperature dependent quadrupolar susceptibility.
From top to bottom, $\alpha$ is equal to 0, 0.04, 0.1, 0.2, 0.4, and 1.
}
\end{figure}
\begin{figure}
\begin{center}
\psfig{figure=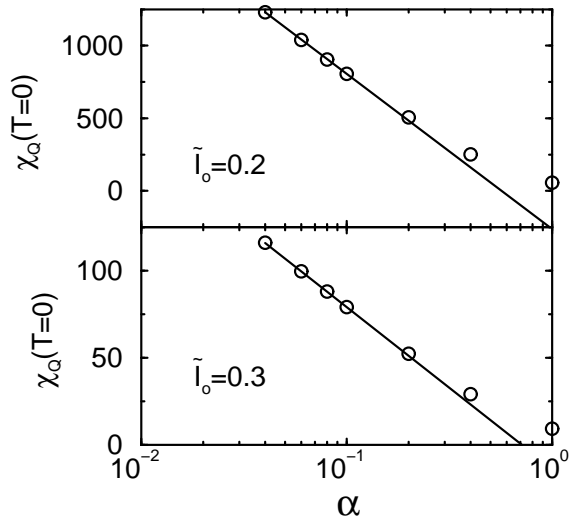,width=7.5cm}
\end{center}
\vspace*{0.05cm}
\caption{
$T=0$ values of quadrupolar susceptibility.
For the small values of the coupling ratio $\alpha$ ($< 0.2$), $\chi_{\rm Q}$
is logarithmic in $\alpha$.
}
\end{figure}
\begin{figure}
\begin{center}
\psfig{figure=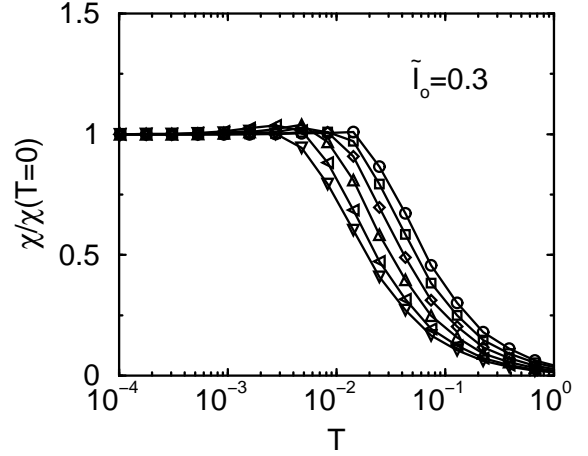,width=7.5cm}
\end{center}
\vspace*{0.05cm}
\caption{
Temperature dependent magnetic susceptibility divided by each $T=0$ value.
From left to right, $\alpha$ is equal to 0, 0.2, 0.4, 0.6, 0.8, and 1.
}
\end{figure}
\begin{figure}
\begin{center}
\psfig{figure=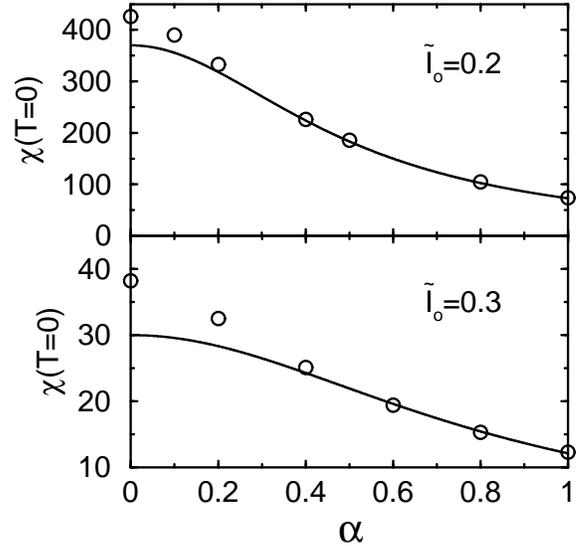,width=7.5cm}
\end{center}
\vspace*{0.05cm}
\caption{
$T=0$ values of magnetic susceptibility.
When the coupling ratio $\alpha$ is not very small, $\chi$ is fitted to
$(A + B \alpha^2)^{-1}$ (solid line), where
$(A,B) = (2.7 \times 10^{-3},0.011)$ for $\tilde{I}_{\rm o} = 0.2$ and
$(3.3 \times 10^{-3},0.049)$ for $\tilde{I}_{\rm o} = 0.3$.
}
\end{figure}

\begin{thebibliography}{99}

\bibitem{Cox98}
For a review see D. L. Cox and A. Zawadowski,
Adv. Phys. {\bf 47}, 599 (1998) and references therein.

\bibitem{Hewson93}
A. C. Hewson,
{\it The Kondo Problem to Heavy Fermions}
(Cambridge University Press, 1993).

\bibitem{Yamada92}
K. Yamada, K. Hanzawa, and K. Yosida,
Prog. Theor. Phys. Suppl. {\bf 108}, 141 (1992).

\bibitem{Ami94}
H. Amitsuka and T. Sakakibara,
J. Phys Soc. Jpn {\bf 63}, 736 (1994).

\bibitem{Nishi99}
S. Nishigori, K. Fujiwara, and T. Ito,
Physica (Amsterdam) {\bf B259-261}, 397 (1999).

\bibitem{Cox93}
D. L. Cox,
Physica (Amsterdam) {\bf B186-188}, 312 (1993).

\bibitem{Aron97}
M. C. Aronson, M. B. Maple, P. de Sa, A. M. Tsvelik, and R. Osborn,
Europhys. Lett. {\bf 40}, 245 (1997).

\bibitem{Andra98}
M. C. de Andrade, R. Chau, R. P. Dickey, N. R. Dilley, E. J. Freeman,
D. A. Gajewski, M. B. Maple, R. Movshovich, A. H. Castro Neto, G. Castilla,
and B. A. Jones,
Phys. Rev. Lett. {\bf 81}, 5620 (1998).

\bibitem{Koga99}
M. Koga, G. Zar{\'a}nd, and D. L. Cox,
submitted to Phys. Rev. Lett. (cond-mat/9903311).

\bibitem{Bick87}
N. E. Bickers, D. L. Cox, and J. W. Wilkins,
Physical Review B {\bf 36}, 2036 (1987).

\bibitem{Hirst77}
L. L. Hirst,
Adv. Phys. {\bf 27}, 231 (1978).

\bibitem{Pava97}
E. Pavarini and L. C. Andreani,
Phys. Rev. B {\bf 56}, 5073 (1997).

\bibitem{Kusu99}
H. Kusunose and Y. Kuramoto,
to be published in J. Phys. Soc. Jpn (cond-mat/9904292).

\bibitem{Hiroi97}
M. Hiroi, S. Kobayashi, M. Sera, N. Kobayashi, and S. Kunii,
J. Phys. Soc. Jpn {\bf 66}, 1762 (1997).

\bibitem{Tayama97}
T. Tayama, T. Sakakibara, K. Tenya, H. Amitsuka, and S. Kunii,
J. Phys. Soc. Jpn {\bf 66}, 2268 (1997).

\bibitem{Schrie66}
J. R. Schrieffer and P. A. Wolff,
Phys. Rev. {\bf 149}, 491 (1966).

\bibitem{Wilson75}
K. G. Wilson,
Rev. Mod. Phys. {\bf 47}, 773 (1975);
O. sakai, Y. Shimizu, and T. Kasuya,
Prog. Theor. Phys. Suppl. {\bf 108}, 73 (1992).

\bibitem{Affl92}
I. Affleck, A. W. W. Ludwig, H.-B. Pang, and D. L. Cox,
Phys. Rev. B {\bf 45}, 7918 (1992).

\bibitem{Lea62}
K. K. Lea, M. J. M. Leask, and W. P. Wolf,
J. Phys. Chem. Solids {\bf 23}, 1381 (1962).

\bibitem{Onode66}
Y. Onodera and M. Okazaki,
J. Phys. Soc. Jpn {\bf 21}, 2400 (1966).

\bibitem{Kim97}
T.-S. Kim, L. N. Oliveira, and D. L. Cox,
Phys. Rev. B {\bf 55}, 12460 (1997).

\bibitem{Affl90}
I Affleck,
Nucl. Phys. B {\bf 336}, 517 (1990);
I. Affleck and A. W. W. Ludwig,
{\it ibid.} {\bf 352}, 849 (1991).

\bibitem{Fabri95}
M. Fabrizio, A. O. Gogolin, and P. Nozi{\`e}res,
Phys. Rev. Lett. {\bf 74}, 4503 (1995).

\bibitem{Andrei95}
N. Andrei and A. Jerez,
Phys. Rev. Lett. {\bf 74}, 4507 (1995).

\bibitem{Sato84}
N. Sato, S. Kunii, I. Oguro, T. Komatsubara, and T. Kasuya,
J. Phys. Soc. Jpn {\bf 53}, 3967 (1984).

\bibitem{Sato85}
N. Sato, A. Sumiyama, S. Kunii, H. Nagano, and T. Kasuya,
J. Phys. Soc. Jpn {\bf 54}, 1923 (1985).

\end{thebibliography}
\end{document}